\documentclass{iau}
\usepackage{natbib}

\title[Understanding PSR~B1828-11]
      {Advances in our understanding of the free precession candidate PSR~B1828-11}

\author[G.~Ashton, D.~I.~Jones, and R. Prix]
{G.~Ashton$^1$ \and D.~I.~Jones$^2$ \and R. Prix$^{1}$}

\affiliation{$^1$ Max Planck Institut f{\"u}r Gravitationsphysik
             (Albert Einstein Institut) and Leibniz Universit\"at Hannover,
             30161 Hannover, DE
             \\ email: {\tt gregory.ashton@ligo.org}
             \\[\affilskip]
             $^2$ Mathematical Sciences,
             University of Southampton,
             Southampton SO17 1BJ, UK
}

\pubyear{2017}
\volume{337}
\jname{Pulsar Astrophysics -­ The Next 50 Years}
\editors{P. Weltevrede, B.B.P. Perera, L. Levin Preston \& S. Sanidas, eds.}

\begin{document}

\maketitle

\begin{abstract}

We highlight the advances and difficulties in understanding PSR~B1828-11, which
undergoes long-term periodic modulations in its timing and pulse shape over
several years. A model comparison of precession and magnetospheric switching
models based on the long-term modulation data favours the former; we discuss
the implications of this in the context of short timescale switching observed
in this pulsar. Furthermore, we highlight the difficulties this pulsar poses
for our understanding of pulsars due to the increasing rate of the modulation
period and its behaviour during a recent glitch.

\end{abstract}

The periodic modulations of PSR~B1828-11 were first tentatively interpreted as
evidence for planetary companions \citep{bailes1993}, before observations of
modulations in the pulse shape, correlated with the timing features, led to its
identification as a free-precession candidate \citep{stairs2000}. Motivated by
mode-nulling and mode-changing events in other pulsars and using data
time-averaged on short timescales, \citet{lyne2010} reinterpreted the
correlation in timing and pulse properties of PSR~B1828-11 (along with those of
several other pulsars) as evidence for magnetospheric switching, a process
whereby the magnetosphere periodically switches between two stable
configurations; the key evidence being that the pulse shape parameters did not
smoothly vary between two extremes (as expected for precession with a
Gaussian core emission \citep{akgun2006}), but instead ``spend most of the time
in just one extreme state or the other''.

Understanding the cause of periodic modulations of PSR~B1828-11 may have
implications for the interior superfluid and may also provide a crucial insight
into understanding the causes of the timing noise experienced by all pulsars at
some level \citep{hobbs2010}.

\emph{Short-term switching} --
When observed over a short duration ($\sim 1$~hr), the pulsar switches between
a wide and a narrow profile, and the proportion of time spent in each profile
varies over the precession cycle, ultimately leading to the observed long-term
variations in the pulse profile \citep{stairs2003}. The short timescale nature
of these switches does suggest they are magnetospheric in origin. But, this
does not rule out a precession model. First, emission profiles with both
core and conal blobs can explain the short-term switching under a precession
interpretation, without any magnetospheric switching \citep{akgun2006}. Second,
precession may act as the clock of switching: the switching being biased by
angles which are periodically varied \citep{jones2012, kerr2016periodic}; see
also the idea of stochastic resonance \citep{cordes2013}.

\emph{Long-term switching} --
A simple question is ``which of the two models, precession or magnetospheric
switching, better explain the long-term timing and pulse shape modulations of
PSR~B1828-11?''. To answer this, we performed a model comparison between
precession and a phenomenological switching model \citep{perera2015}; to make
the comparison fair, we conditioned each model on the spin-down data and then
calculated a Bayes factor between the two using the pulse shape data; the Bayes
factor was found to favour the precession interpretation by a factor of
$10^{2.7\pm 0.5}$ \citep{ashton2016}. On the basis of the long-term switching
alone, precession is far from ruled out, but rather favoured over this
switching model. This conclusion requires there is a mechanism similar to that
proposed by \citet{akgun2006} to explain the short-term changes in the pulse
profile: further
study understanding the plausibility of such a mechanism may therefore prove
useful.

\emph{Problems with precession} --
We identified two further challenges to understanding PSR~B1828-11.  First, the
$\sim 500$~day modulation period has gradually been getting shorter over the
full observation span at a rate of $\approx -0.01$~s/s; in the context of
precession, this unexpected results suggests that the deformation is growing on
a timescale of $\sim213$~yr~\citep{ashton2017}.  A planetary explanation may
provide a more natural explanation, although it faces challenges in explaining
variations in the pule profile.  Second, on MJD~55042, the pulsar underwent a
glitch \citep{espinoza2011}, but the modulation appears to be unaffected; this
demonstrates inconsistencies in our understanding of precession or even
glitches \citep{jones2017}.

\emph{Conclusions} --
Though PSR~B1828-11 is well behaved compared to most pulsars (in that its
timing anomalies are stable and periodic), no complete model is able to explain
all the features. But, by systematic study, we believe a lot more can be
learnt. Further high-resolution observations will shed light on the interplay
between the short-term switching and the long-term behaviour; extending the
study of the modulation period to a longer set of data after the glitch, one
could test for sudden changes in the modulations during the glitch,
elucidating any dependence on the crust; studies of the polarisation (as done
by \citet{weisberg2010}) could also help us to better understand the system. By
combining all of these observations and comparing predictive models, it seems
very promising that progress can be made on understanding this object.

\vspace{-3mm}
\bibliographystyle{apalike}
\bibliography{bibliography}

\end{document}